\documentclass[sigconf]{acmart}

\usepackage{multirow}
\usepackage{todonotes}
\usepackage{enumitem}

\AtBeginDocument{%
  \providecommand\BibTeX{{%
    \normalfont B\kern-0.5em{\scshape i\kern-0.25em b}\kern-0.8em\TeX}}}

\copyrightyear{2020}
\acmYear{2020}
\setcopyright{acmcopyright}
\acmConference[CIKM '20]{Proceedings of the 29th ACM International Conference on Information and Knowledge Management}{October 19--23, 2020}{Virtual Event, Ireland}
\acmBooktitle{Proceedings of the 29th ACM International Conference on Information and Knowledge Management (CIKM '20), October 19--23, 2020, Virtual Event, Ireland}
\acmPrice{15.00}
\acmDOI{10.1145/3340531.3417428}
\acmISBN{978-1-4503-6859-9/20/10}

\begin{document}

\title{\textsc{CovidExplorer}: A Multi-faceted AI-based Search and Visualization Engine for COVID-19 Information}

\author{Heer Ambavi}
\authornote{These authors have contributed equally.}
\author{Kavita Vaishnaw}
\authornotemark[1]
\author{Udit Vyas}
\authornotemark[1]
\affiliation{%
  \institution{Indian Institute of Technology, Gandhinagar}
  \streetaddress{P.O. Box 1212}
  \city{Gandhinagar}
  \state{Gujarat}
  \country{India}
  \postcode{382355}
}

\author{Abhisht Tiwari}
\author{Mayank Singh}
\authornote{Corresponding author can be contacted at: singh.mayank@iitgn.ac.in}
\affiliation{%
  \institution{Indian Institute of Technology, Gandhinagar}
  \streetaddress{P.O. Box 1212}
  \city{Gandhinagar}
  \state{Gujarat}
  \country{India}
  \postcode{382355}
}

\renewcommand{\shortauthors}{Ambavi, et al.}

\begin{abstract}
  The entire world is engulfed in the fight against the COVID-19 pandemic, leading to a significant surge in research experiments, government policies, and social media discussions. A multi-modal information access and data visualization platform can play a critical role in supporting research aimed at understanding and developing preventive measures for the pandemic. In this paper, we present a multi-faceted AI-based search and visualization engine, \textsc{CovidExplorer}. Our system aims to help researchers understand current state-of-the-art COVID-19 research, identify research articles relevant to their domain, and visualize real-time trends and statistics of COVID-19 cases. In contrast to other existing systems, \textsc{CovidExplorer} also brings in India-specific topical discussions on social media to study different aspects of COVID-19. The system, demo video, and the datasets are available at \url{http://covidexplorer.in}.
\end{abstract}




\maketitle

\section{Introduction}
With an exponential growth rate in COVID-19 infections, all government and private organizations are spending heavily on R\&D infrastructure and essential medical facilities. This has led to a high surge in scientific volume ranging from proposal for innovative medical devices, vaccines, infection prediction, and propagation models for COVID-19. For instance, we witness several tongue-in-cheek media headlines like \textit{`Scientists are drowning in COVID-19 papers. Can new tools keep them afloat?'}\footnote{https://www.sciencemag.org/news/2020/05/scientists-are-drowning-covid-19-papers-can-new-tools-keep-them-afloat}. Several COVID-19 specific search and recommendation tools have been developed recently by various research groups.\footnote{\url{https://discourse.cord-19.semanticscholar.org/t/cord-19-demos-and-resources/132}} However, to the best of our knowledge, systems that leverage graph-based interactive visualizations to navigate the vast research volume are not available. 

A large volume of social media discussions around the COVID-19 has also led to several natural language processing (NLP) limitations like identifying facts or opinionated messages, COVID-19 specific trending topics, and fake or hate messages. Except for COVID-specific data curation strategies~\cite{chen2020covid19}, we find a few works~\cite{hosseini2020content,muller2020covid} that address some of these NLP limitations. None of the available systems has attempted to encapsulate social media discussions with scholarly search and recommendation. 

In this paper, we propose a multi-faceted AI-based search and visualization engine, \textsc{CovidExplorer}. Our system aims to serve as a means for researchers to understand COVID-19 research and visualize the trends in the expanding pool of scientific articles on coronaviruses. Our system seamlessly integrates three different aspects of COVID-19 into a single platform (i) search and recommendation, (ii) statistics, and (iii) social media discussions. The system aims to facilitate the researchers and decision-makers with the latest global research updates and social discussions in India.

\section{Dataset}
\label{sec:dataset}
The development of \textsc{CovidExplorer} leverages two rich time-stamped datasets. The first dataset (hereafter \textbf{`CORD-19'} dataset) comprises $\sim$157,000 scholarly articles, including over 75,000 full-text articles on coronaviruses. 

\begin{table}[h]
\centering
  \begin{tabular}{lll} \toprule
   \parbox[t]{2mm}{\multirow{3}{*}{\rotatebox[origin=c]{90}{\textbf{\small CORD-19}}}}
   &Year range& 1991--2020 \\ 
  &Number of papers &157,712 \\
  &Open access papers& 119,862 \\\hline
  \parbox[t]{2mm}{\multirow{3}{*}{\rotatebox[origin=c]{90}{\textbf{Tweet}}}}
   &Number of tweets &226,989,304\\ 
  &Month range& Jan--May, 2020 \\ 
  &Total Indian Tweets& 5,787,974\\
  \bottomrule
 \end{tabular}
 \caption{General statistics about the \textit{CORD-19} publication and \textit{Tweet dataset} (As of 18 June 2020).}\label{tab:dataset}
 \vspace{-5mm}
\end{table}

\begin{figure*}[h!]
\begin{tabular}{@{}c@{}c@{}c@{}}
  \includegraphics[width=0.50\hsize]{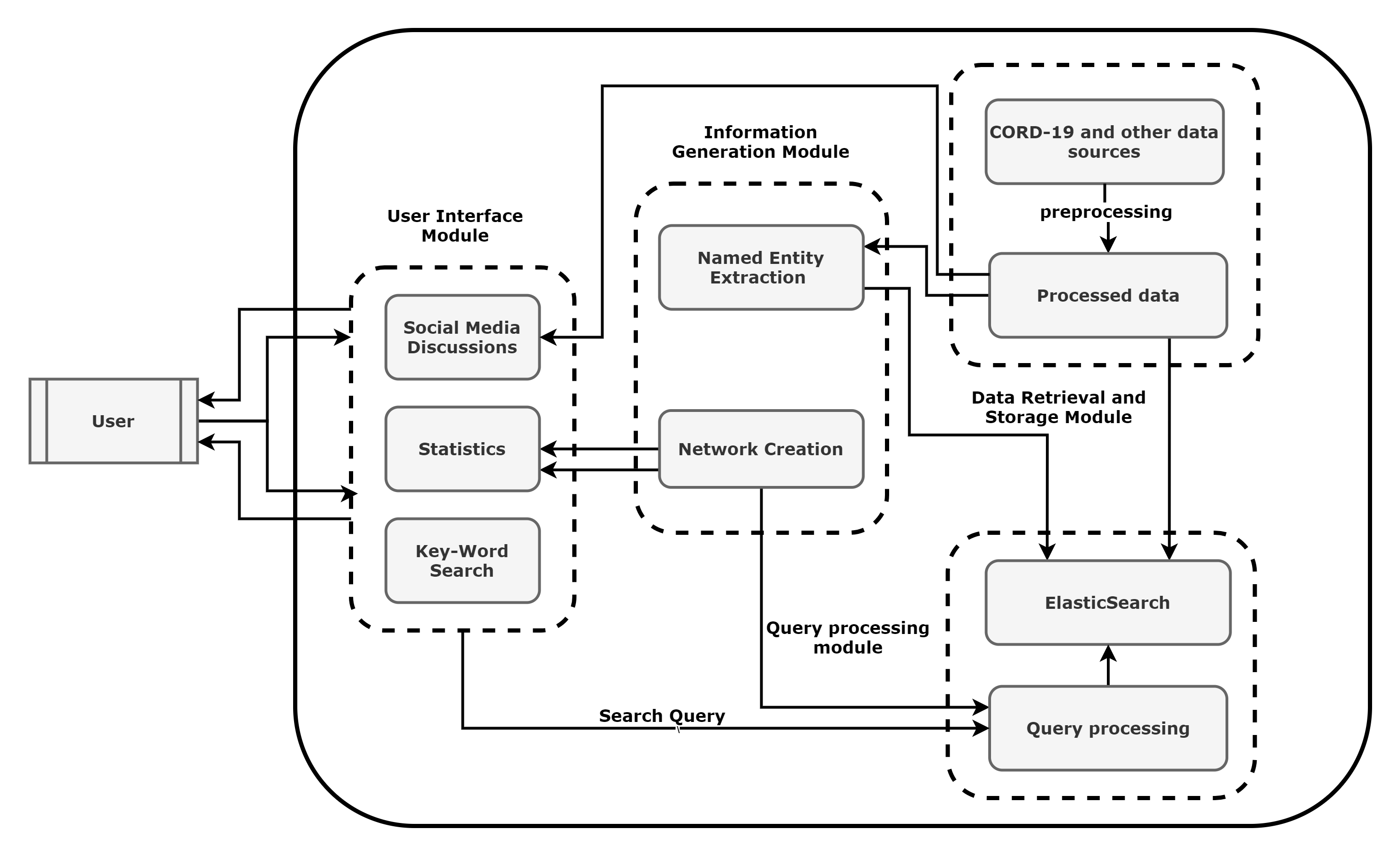} &&
  \includegraphics[width=0.50\hsize]{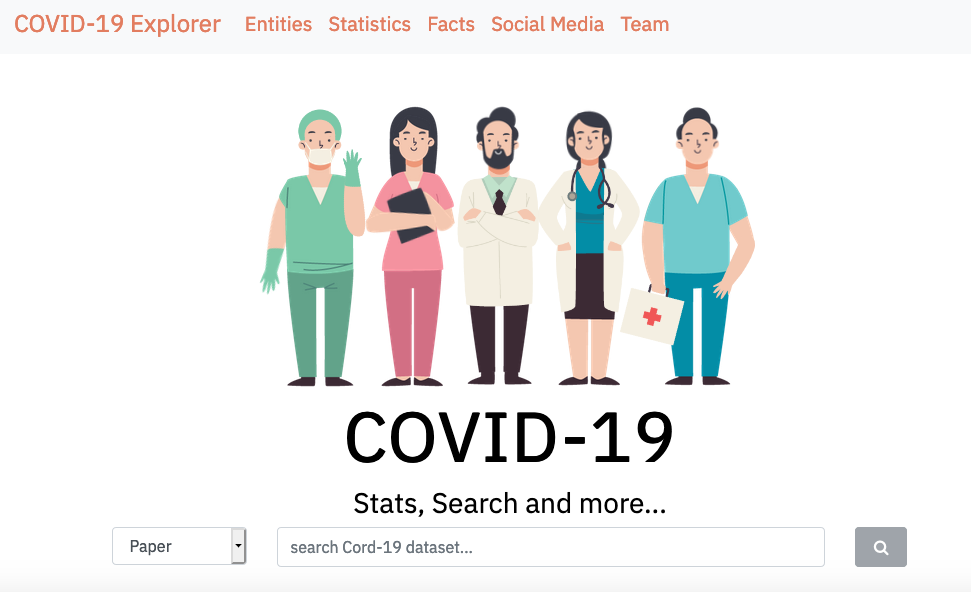} \\
  (a) & &(b) \\
\end{tabular}
\vspace{-2mm}
\caption{(a)~Architecture of \textsc{CovidExplorer}. (b)~The landing page.}  
\label{fig:architecture}
\vspace{-2mm}
\end{figure*}
The White House and a coalition of leading research groups provide CORD-19~\cite{Wang2020CORD19TC} to the global research community. We also curate about 227 million tweets relevant to the COVID-19 pandemic (hereafter \textbf{`Tweet'} dataset). We periodically collect relevant tweet IDs from a publicly available COVID-19 TweetID corpus~\cite{chen2020covid19}. The tweet IDs are hydrated using the DocNow Hydrator tool\footnote{\url{https://github.com/docnow/hydrator}}. We only explore India-specific tweets by using location metadata or the presence of Indian locations (states, cities, towns, and village names) in the tweet text. We also include tweets that are either posted as a reply to the India-specific tweets and tweets for which India-specific tweets were posted as a reply.  
In this paper, we present all our analyses on this subset of tweets. Both datasets are updated weekly. Table~\ref{tab:dataset} presents statistics of the \textit{CORD-19} and Tweet dataset. We release the processed data with a clear and accessible data usage license\footnote{https://creativecommons.org/licenses/by/4.0/legalcode}.
\vspace{-4mm}
\section{Architecture}
In this section, we present the detailed architecture of \textsc{CovidExplorer}. The proposed system comprises four interdependent modules: (i) Data Retrieval and Storage, (ii) Information Generation, (iii) Query Processing, and (iv) User Interface. Figure~\ref{fig:architecture}a shows the system architecture. Figure~\ref{fig:architecture}b shows the landing page of \textsc{CovidExplorer}. 

\begin{enumerate}[noitemsep,nolistsep]
    \item \textbf{Data Retrieval and Storage Module}: This module periodically downloads the data updates from both the data sources (described in Section~\ref{sec:dataset}) and performs pre-processing.
    The pre-processed data is stored in a semi-structured format to facilitate a quick query response. Elasticsearch engine indexes the \textit{CORD-19 dataset}, and the \textit{Tweet dataset} is stored as CSV dumps. 
    
    \item \textbf{Information Generation Module}: This module processes the curated data. It extracts biological entities and updates entity statistics (Section~\ref{sec:entity}) from the \textit{CORD-19 dataset}. The Elasticsearch engine also indexes this generated entity information. In the case of \textit{Tweet data}, it performs basic location-based filtering, generates timelines, performs topic modeling, and generates India-specific insights.
    
    \item \textbf{Query Processing Module}: This module is responsible for answering user queries. It fetches query results from the processed data. For \textit{CORD-19 data}, this module supports the keyword search, timeline, and entity queries. It also processes temporal queries about tweet data.
    
\item \textbf{User Interface Module}: The User Interface module renders the query results using interactive visualizations.\\
\end{enumerate}

We use Flask\footnote{\url{https://palletsprojects.com/p/flask/}} framework,  written in Python, for web deployment. We use Elasticsearch\footnote{\url{https://www.elastic.co/elasticsearch/}} for data storage and querying. The timelines are generated using TimelineJS\footnote{\url{http://timeline.knightlab.com/}} and plots are rendered using Plotly\footnote{\url{https://plot.ly/javascript}}, Dash\footnote{\url{https://plotly.com/dash/}} and amCharts\footnote{\url{https://www.amcharts.com/}}.
The tweet preprocessing is done using NLTK and Gensim libraries.
\begin{figure}[h]
\centering
\includegraphics[scale = 0.2]{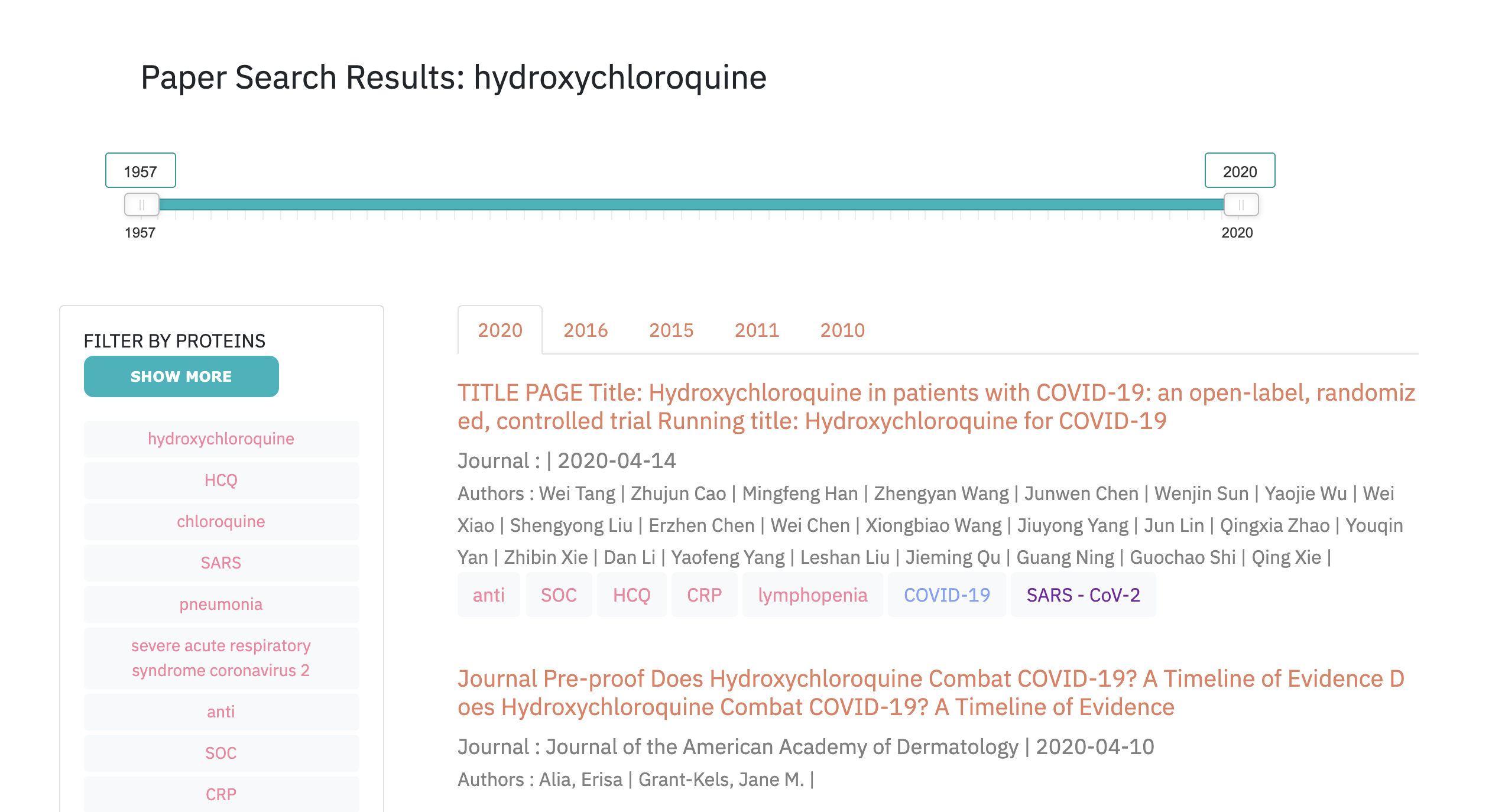}
\setlength{\intextsep}{-15pt}
\caption {A representative query \textit{`hydroxychloroquine'} results in a list of relevant papers distributed year-wise. Each search result is displayed along with venue name, date of publication, authors, bio-entity mentions and URL to the original source.}
\label{papersearch}
\vspace{-4mm}
\end{figure}

\begin{figure*}[h]
\centering
\small{
\begin{tabular}{cc}
    \includegraphics[scale = 0.1]{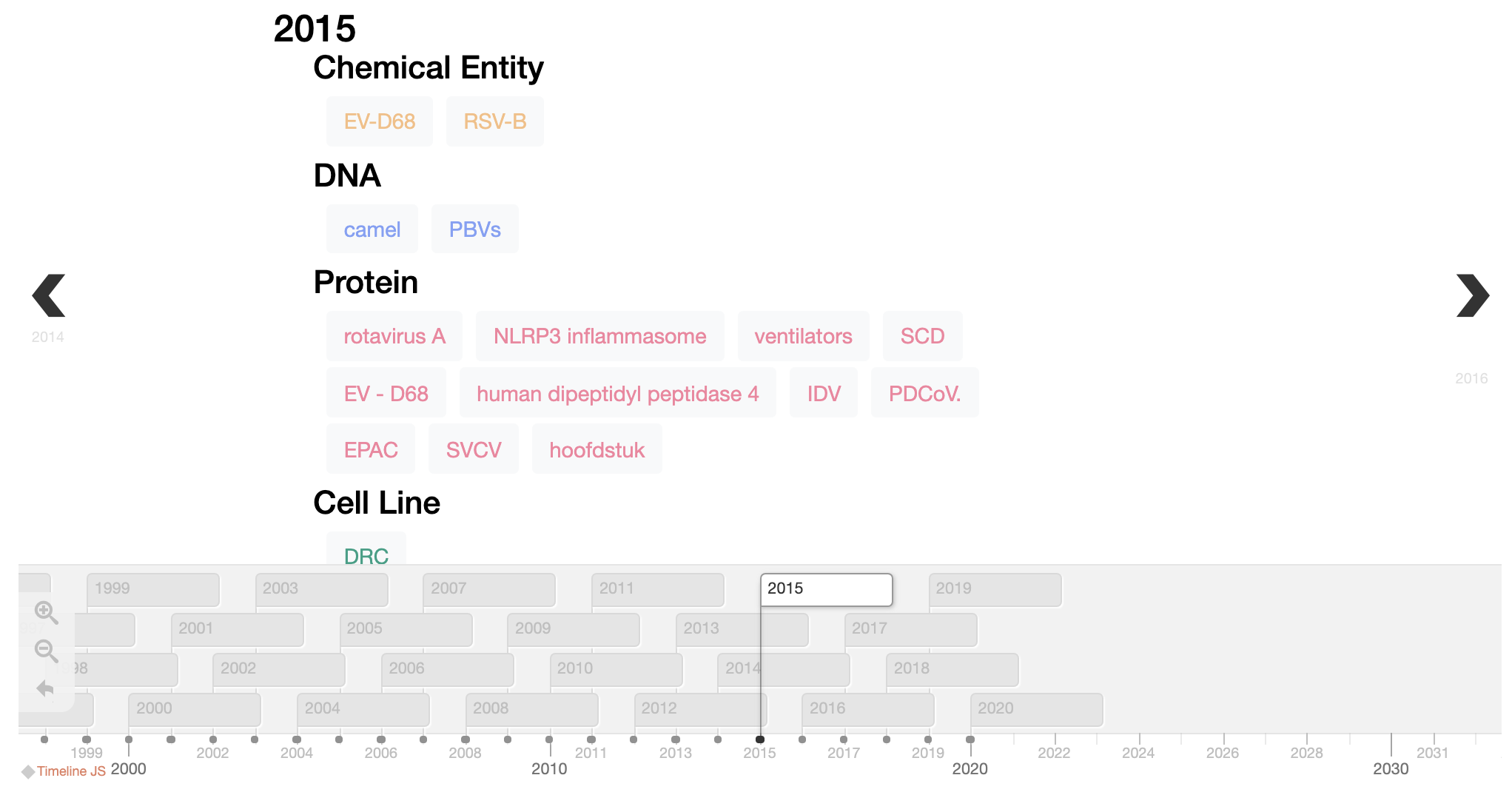} & \includegraphics[scale = 0.18]{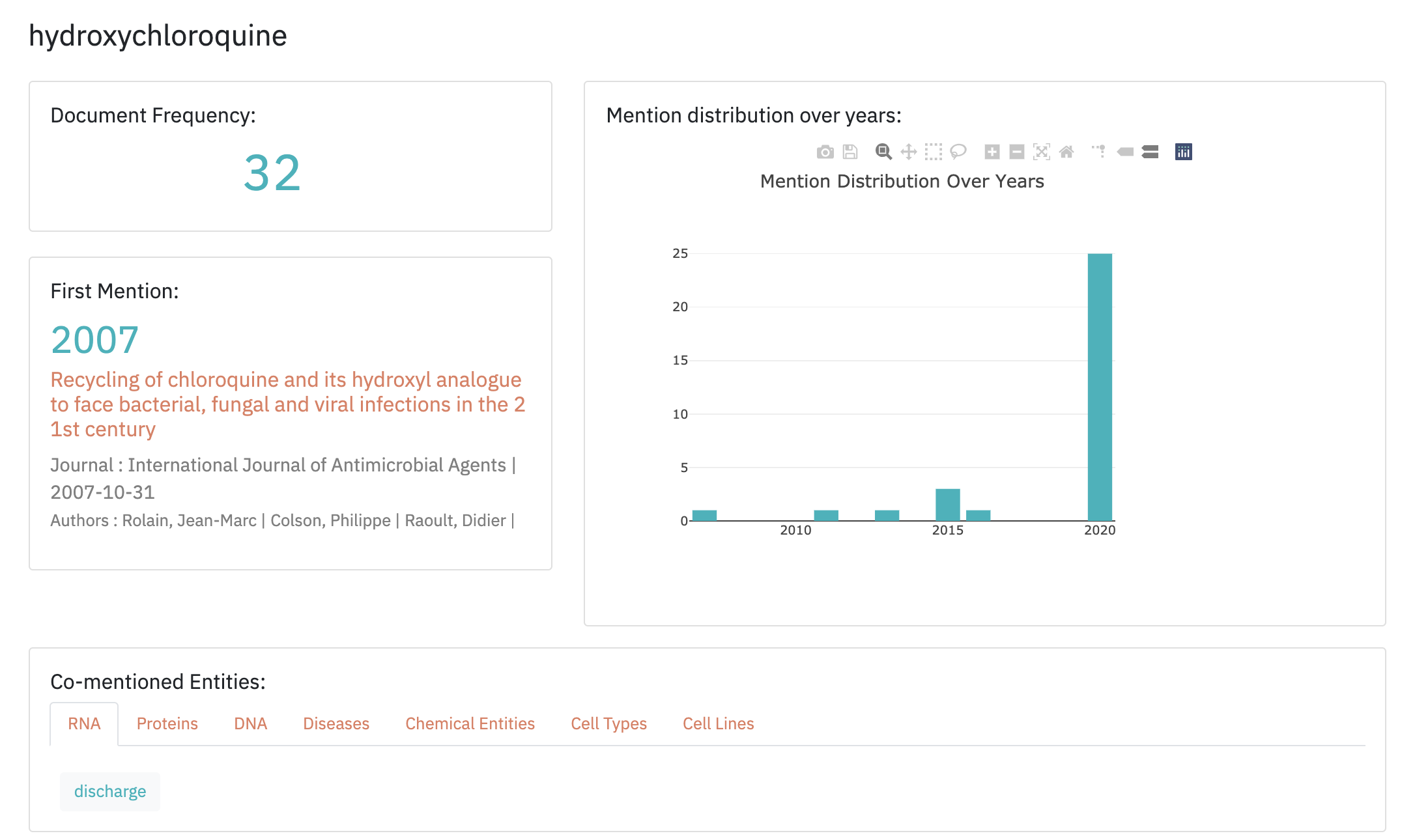} \\
     (a)& (b) \\
     \multicolumn{2}{c}{\includegraphics[scale = 0.2]{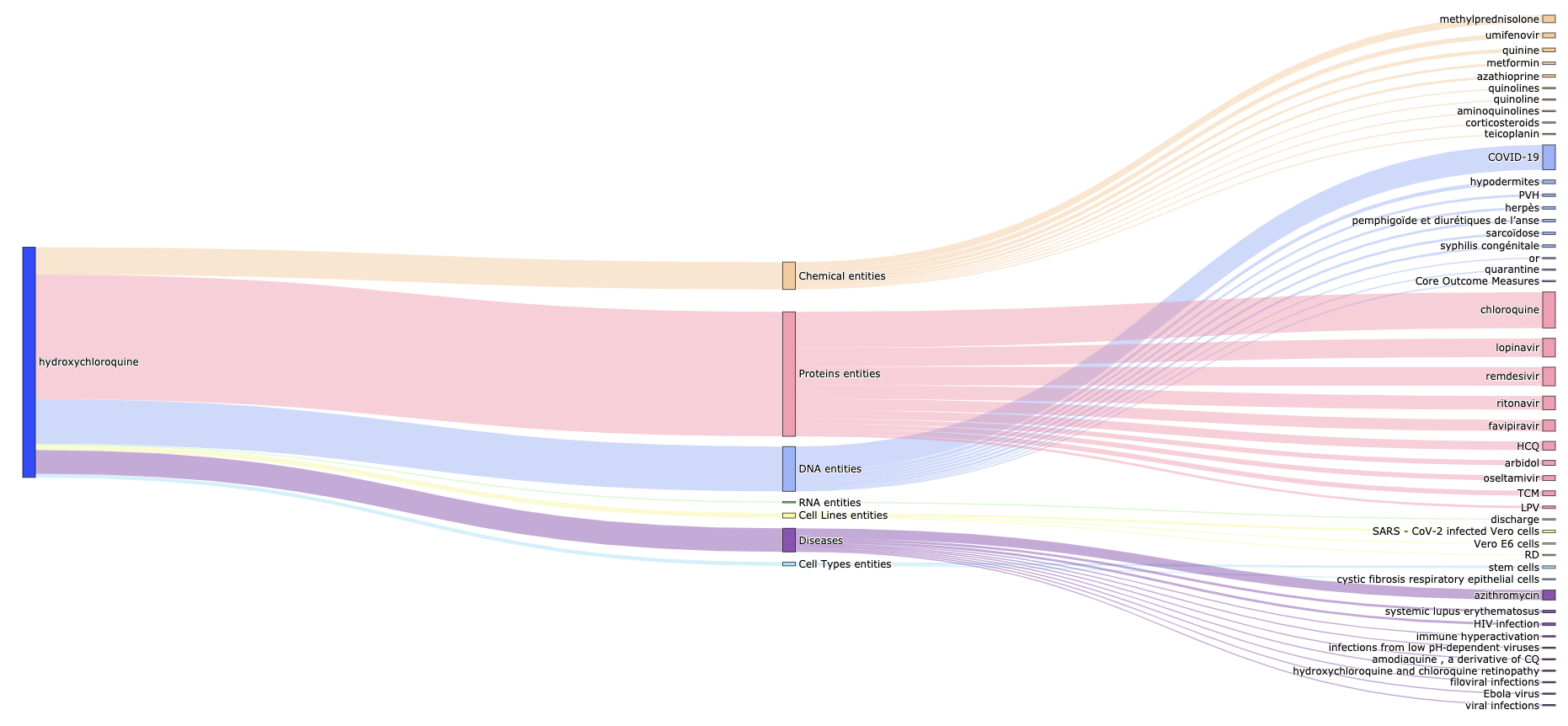}}\\ 
     \multicolumn{2}{c}{(c)}\\
\end{tabular}}
    \vspace{-4mm}
    \caption{Visualizing biological entities. (a)~ Timeline of first mention of an entity instance segregated into different entity types. (b)~Statistics page for entity \textit{ `Hydroxychloroquine'}. (c)~Top-10 co-mentions for all entity types for entity \textit{ Hydroxychloroquine}. The flow values between \textit{Hydroxychloroquine} and the entity type is the total number of co-mentions of that type.}
    \label{fig:entity}
    \vspace{-3mm}
\end{figure*}

\section{Functionalities}

\subsection{COVID-19 Scholarly Search}
We develop COVID-19 scholarly search facilities using the \textit{CORD-19 dataset}. The search facility comprises several components described in the following sections. 

\subsubsection{Full-Text Search}
\textsc{CovidExplorer} supports a keyword-based search in three categories: authors, papers (title), and full-text papers. In addition to the text of the articles, metadata such as title, abstract, author names, publication year, and venue is used. Each search query list of relevant papers displayed along with their authors, venue, the date of publication and bio-entities mention (described in the next section), and segregated by the year of publication. Each search result is linked to the original paper source URL. These search results can be further filtered by applying a range of publication year, or by their bio-entity mentions. Figure~\ref{papersearch} shows one example of paper search results for keyword \textit{`hydroxychloroquine'}.

\subsubsection{Named Entity Recognition and Visualization}
\label{sec:entity}
\textsc{CovidExplorer} is equipped with a Named Entity Recognition (NER) system for aiding navigation through the large volume of papers. The NER system uses the state-of-the-art language model for scientific and biomedical text SciBERT~\cite{Beltagy2019SciBERT}. We fine-tune SciBERT using the JNLPBA corpus~\cite{DBLP:journals/corr/abs-1901-10219} and the NCBI-disease corpus~\cite{NCBI}. Every article in the search result shows the biological entities extracted from its abstract. The current NER functionality identifies seven different types of bio-entities: DNA, RNA, proteins, cell types, cell lines, diseases, and chemical names. Any entity belonging to multiple types is assigned the maximally occurring type for that entity. We assume that other entity types (except disease) are sub-types of the chemical name entity type. Hence, other types are given priority over the chemical names type during the assignment.

\begin{table*}[ht]
    \centering
    \begin{tabular}{lccl}
    \toprule
        \textbf{Name}&\textbf{Frequency}&\textbf{Unique Entities}&\textbf{Top-5 mentions}\\\hline
        DNA & 68,169 & 33,392 & COVID-19, nucleotide, viral genome, 2019-nCoV, open reading frames \\
        RNA & 11,487 & 4,549 & viral RNA, mRNA, miRNAs, SARS-CoV-2 RNA, lncRNAs \\
        Proteins& 169,160 & 57,956 & SARS, SARS-CoV-2, coronavirus, influenza, pneumonia\\
        Cell Types& 25,617 & 7,632 & macrophages, host cells, infected cells, T cells, HCWs\\
        Cell Lines& 11,357 & 5,662 & Vero cells, cell lines, cell culture, Vero E6 cells, A549 cells\\
        Diseases& 240,956 & 56,364 & coronavirus disease, infectious disease, severe acute respiratory \\
        & & & syndrome, viral infections, MERS-CoV\\
        Chemical Names& 39,331 &10,661 & amino acid, nucleotides, serine, tuberculose, infectieziekten\\\hline
    \end{tabular}
    \caption{Total number of instances along with top-5 mentions in each entity class.}
    \label{tab:entity}
    \vspace{-7mm}
\end{table*}

We provide multiple insights about the recognized entities. For each entity type, we display a timeline to visualize the first mention of each entity of that type, shown in Figure \ref{fig:entity}a. We also list the top-10 most frequently mentioned entities for each entity type. Table~\ref{tab:entity} shows the statistics along with the names of popular instances in each entity class. For each entity, we also individually show the first mention, a visualization of popular co-mentioned entities, and the year-wise distribution of mention frequencies. We also display all papers that contain these individual entities segregated by the year of publication. Figures~\ref{fig:entity}b and~\ref{fig:entity}c show the entity statistics and co-mentioned entities for candidate entity \textit{`Hydroxychloroquine'}.
\vspace{-2mm}
\subsection{India-specific Infection Statistics}
\textsc{CovidExplorer} also displays a statistics page that keeps track of the daily evolving pandemic situation in India. Daily, it updates the total cases, active cases, deaths, and recovery counts. Besides, it provides state-wise cumulative data of the number of active cases. We fetch the entire infection statistics from the official portal of the Ministry of Health and Family Welfare (MoHFW, India)\footnote{\url{https://www.mohfw.gov.in/}}.
\begin{table}[t]
    \centering
    \label{LQMS}
    \begin{tabular}{lccl}
    \toprule
    \textbf{Month}&\textbf{Number of URLs}&\textbf{LQMS \%}\\\hline
    Jan & 93,252 & 2.04\%\\
    Feb & 212,523 & 1.89\%\\
    Mar & 307,008 & 0.81\%\\
    Apr & 376,027 & 0.70\%\\
    May & 365,165 & 0.77\%\\\hline
    \end{tabular}
    \caption{Presence of Low Quality Misinformation Sources (LQMS) in tweet volumes.}
    \label{tab:LQMS}
    \vspace{-5mm}
\end{table}

\subsection{Social Media Discussions}

The filtered dataset of $\sim$5.7 million tweets is processed and in the following sections, we describe some of the India-specific social media insights.
\begin{figure}[!tbh]
\label{socmed}
\centering
\includegraphics[scale = 0.22]{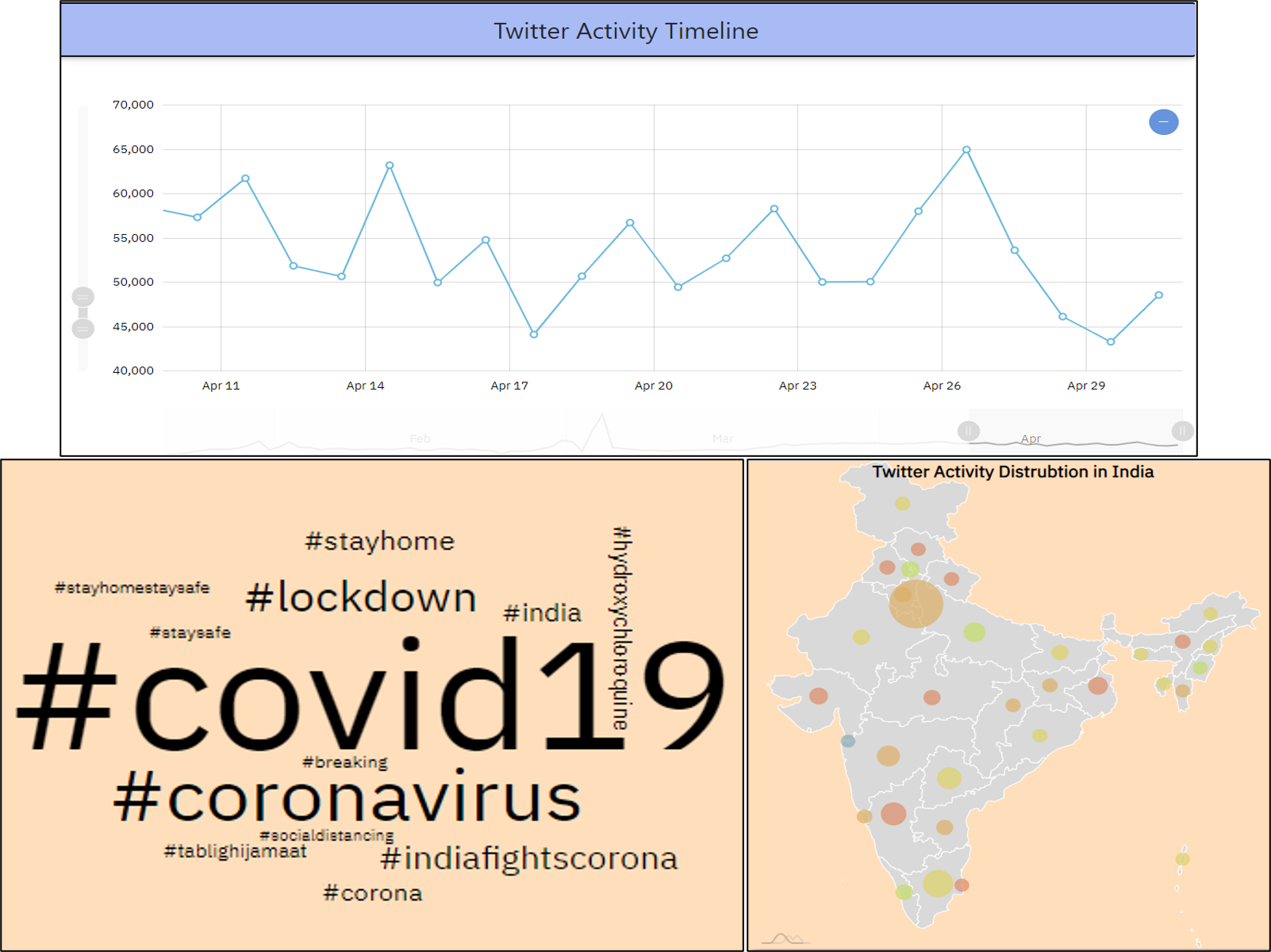}
\caption{Activity timeline, common hashtags and geographical distribution of tweets across India.}
\label{fig:social_media}
\vspace{-5mm}
\end{figure}

\subsubsection{Temporal Activity and Trends}
\textsc{CovidExplorer} Social media page displays an interactive timeline of the Twitter activity through the duration of the pandemic, as well as a state-wise geographical distribution of the twitter activity in India. It also shows most common hashtags, mentions, tweet locations, and URLs. The shared URLs are examined to determine the most common domain names tweeted. To quantify the misinformation prevailing in the tweet volumes, we classify the mentioned domains into misinformation sources and trusted source categories. We use the list of \textit{low-quality misinformation sources (LQMS) } curated by NewsGuard\footnote{\url{https://www.newsguardtech.com/coronavirus-misinformation-tracking-center/}}. Table~\ref{tab:LQMS} shows the statistics of the presence of LQMS in the \textit{Tweet dataset}. Figure~\ref{fig:social_media} shows visualisations of social media interactions.

\subsubsection{Temporal Topic Evolution} The tweets are further processed using the Latent Dirichlet Allocation (LDA) algorithm to infer topical distribution. The LDA results are interactively displayed with the month-wise distribution of topics and the keywords for each topic along with their probabilities. 

\section{Conclusion and Future Directions}
In this work, we present \textsc{CovidExplorer}. \textsc{CovidExplorer} aids researchers in understanding current state-of-the-art COVID-19 research, identify research articles relevant to their domain, visualize real-time trends and statistics of COVID-19 cases, and understand social media discussions. In the future, we aim to provide APIs to enable researchers to access the query results. We also aim to create named entity-based networks that visually depict the context of occurrences of the entities across the resource pool. The \textit{Tweet dataset} can be further processed using BERT-like models to construct sentence embeddings for unsupervised topical clustering.  This method can be an effective method of labeling the dataset. 

\begin{acks}
This work was partially supported by Google Cloud COVID-19 credits program.
\end{acks}
\bibliographystyle{ACM-Reference-Format}
\bibliography{base}


\begin{thebibliography}{7}


\ifx \showCODEN    \undefined \def \showCODEN     #1{\unskip}     \fi
\ifx \showDOI      \undefined \def \showDOI       #1{#1}\fi
\ifx \showISBNx    \undefined \def \showISBNx     #1{\unskip}     \fi
\ifx \showISBNxiii \undefined \def \showISBNxiii  #1{\unskip}     \fi
\ifx \showISSN     \undefined \def \showISSN      #1{\unskip}     \fi
\ifx \showLCCN     \undefined \def \showLCCN      #1{\unskip}     \fi
\ifx \shownote     \undefined \def \shownote      #1{#1}          \fi
\ifx \showarticletitle \undefined \def \showarticletitle #1{#1}   \fi
\ifx \showURL      \undefined \def \showURL       {\relax}        \fi
\providecommand\bibfield[2]{#2}
\providecommand\bibinfo[2]{#2}
\providecommand\natexlab[1]{#1}
\providecommand\showeprint[2][]{arXiv:#2}

\bibitem[\protect\citeauthoryear{Beltagy, Lo, and Cohan}{Beltagy
  et~al\mbox{.}}{2019}]%
        {Beltagy2019SciBERT}
\bibfield{author}{\bibinfo{person}{Iz Beltagy}, \bibinfo{person}{Kyle Lo},
  {and} \bibinfo{person}{Arman Cohan}.} \bibinfo{year}{2019}\natexlab{}.
\newblock \showarticletitle{SciBERT: A Pretrained Language Model for Scientific
  Text}. In \bibinfo{booktitle}{\emph{Proceedings of the 2019 Conference on
  Empirical Methods in Natural Language Processing and the 9th International
  Joint Conference on Natural Language Processing (EMNLP-IJCNLP)}}.
  \bibinfo{pages}{3606--3611}.
\newblock


\bibitem[\protect\citeauthoryear{Chen, Lerman, and Ferrara}{Chen
  et~al\mbox{.}}{2020}]%
        {chen2020covid19}
\bibfield{author}{\bibinfo{person}{Emily Chen}, \bibinfo{person}{Kristina
  Lerman}, {and} \bibinfo{person}{Emilio Ferrara}.}
  \bibinfo{year}{2020}\natexlab{}.
\newblock \bibinfo{title}{COVID-19: The First Public Coronavirus Twitter
  Dataset}.
\newblock
\newblock
\showeprint[arxiv]{2003.07372}~[cs.SI]


\bibitem[\protect\citeauthoryear{Do\u{g}an, Leaman, and Lu}{Do\u{g}an
  et~al\mbox{.}}{2014}]%
        {NCBI}
\bibfield{author}{\bibinfo{person}{Rezarta~Islamaj Do\u{g}an},
  \bibinfo{person}{Robert Leaman}, {and} \bibinfo{person}{Zhiyong Lu}.}
  \bibinfo{year}{2014}\natexlab{}.
\newblock \showarticletitle{NCBI Disease Corpus}.
\newblock \bibinfo{journal}{\emph{J. of Biomedical Informatics}}
  \bibinfo{volume}{47}, \bibinfo{number}{C} (\bibinfo{date}{Feb.}
  \bibinfo{year}{2014}), \bibinfo{pages}{1–10}.
\newblock
\showISSN{1532-0464}


\bibitem[\protect\citeauthoryear{Hosseini, Hosseini, and Broniatowski}{Hosseini
  et~al\mbox{.}}{2020}]%
        {hosseini2020content}
\bibfield{author}{\bibinfo{person}{Pedram Hosseini}, \bibinfo{person}{Poorya
  Hosseini}, {and} \bibinfo{person}{David~A Broniatowski}.}
  \bibinfo{year}{2020}\natexlab{}.
\newblock \showarticletitle{Content analysis of Persian/Farsi Tweets during
  COVID-19 pandemic in Iran using NLP}.
\newblock \bibinfo{journal}{\emph{arXiv preprint arXiv:2005.08400}}
  (\bibinfo{year}{2020}).
\newblock


\bibitem[\protect\citeauthoryear{Huang, Lai, Tsai, and Hsu}{Huang
  et~al\mbox{.}}{2019}]%
        {DBLP:journals/corr/abs-1901-10219}
\bibfield{author}{\bibinfo{person}{Ming{-}Siang Huang},
  \bibinfo{person}{Po{-}Ting Lai}, \bibinfo{person}{Richard~Tzong{-}Han Tsai},
  {and} \bibinfo{person}{Wen{-}Lian Hsu}.} \bibinfo{year}{2019}\natexlab{}.
\newblock \showarticletitle{Revised {JNLPBA} Corpus: {A} Revised Version of
  Biomedical {NER} Corpus for Relation Extraction Task}.
\newblock \bibinfo{journal}{\emph{CoRR}}  \bibinfo{volume}{abs/1901.10219}
  (\bibinfo{year}{2019}).
\newblock
\showeprint[arxiv]{1901.10219}
\urldef\tempurl%
\url{http://arxiv.org/abs/1901.10219}
\showURL{%
\tempurl}


\bibitem[\protect\citeauthoryear{M{\"u}ller, Salath{\'e}, and
  Kummervold}{M{\"u}ller et~al\mbox{.}}{2020}]%
        {muller2020covid}
\bibfield{author}{\bibinfo{person}{Martin M{\"u}ller}, \bibinfo{person}{Marcel
  Salath{\'e}}, {and} \bibinfo{person}{Per~E Kummervold}.}
  \bibinfo{year}{2020}\natexlab{}.
\newblock \showarticletitle{COVID-Twitter-BERT: A Natural Language Processing
  Model to Analyse COVID-19 Content on Twitter}.
\newblock \bibinfo{journal}{\emph{arXiv preprint arXiv:2005.07503}}
  (\bibinfo{year}{2020}).
\newblock


\bibitem[\protect\citeauthoryear{Wang, Lo, Chandrasekhar, Reas, Yang, Eide,
  Funk, Kinney, Liu, Merrill, Mooney, Murdick, Rishi, Sheehan, Shen, Stilson,
  Wade, Wang, Wilhelm, Xie, Raymond, Weld, Etzioni, and Kohlmeier}{Wang
  et~al\mbox{.}}{2020}]%
        {Wang2020CORD19TC}
\bibfield{author}{\bibinfo{person}{Lucy~Lu Wang}, \bibinfo{person}{Kyle Lo},
  \bibinfo{person}{Yoganand Chandrasekhar}, \bibinfo{person}{Russell Reas},
  \bibinfo{person}{Jiangjiang Yang}, \bibinfo{person}{Darrin Eide},
  \bibinfo{person}{Kathryn Funk}, \bibinfo{person}{Rodney~Michael Kinney},
  \bibinfo{person}{Ziyang Liu}, \bibinfo{person}{William. Merrill},
  \bibinfo{person}{Paul Mooney}, \bibinfo{person}{Dewey~A. Murdick},
  \bibinfo{person}{Devvret Rishi}, \bibinfo{person}{Jerry Sheehan},
  \bibinfo{person}{Zhihong Shen}, \bibinfo{person}{Brandon Stilson},
  \bibinfo{person}{Alex~D. Wade}, \bibinfo{person}{Kuansan Wang},
  \bibinfo{person}{Christopher Wilhelm}, \bibinfo{person}{Boya Xie},
  \bibinfo{person}{Douglas~M. Raymond}, \bibinfo{person}{Daniel~S. Weld},
  \bibinfo{person}{Oren Etzioni}, {and} \bibinfo{person}{Sebastian Kohlmeier}.}
  \bibinfo{year}{2020}\natexlab{}.
\newblock \showarticletitle{CORD-19: The Covid-19 Open Research Dataset}.
\newblock \bibinfo{journal}{\emph{ArXiv}}  \bibinfo{volume}{abs/2004.10706}
  (\bibinfo{year}{2020}).
\newblock


\end{thebibliography}

\end{document}